\documentclass[twocolumn]{aastex631}

\usepackage{amsmath}
\usepackage{amsfonts}
\usepackage{graphicx}
\usepackage{rotating}
\usepackage{natbib}
\usepackage{txfonts}
\usepackage{color}
\usepackage{wrapfig}
\usepackage{amssymb}
\usepackage{color}

\shorttitle{Hot cores in the SMC} 
\shortauthors{T. Shimonishi et al.} 

\begin{document}

\title{The Detection of Hot Molecular Cores in the Small Magellanic Cloud}

\correspondingauthor{Takashi Shimonishi} 
\email{shimonishi@env.sc.niigata-u.ac.jp} 

\author[0000-0002-0095-3624]{Takashi Shimonishi} 
\affiliation{Institute of Science and Technology, Niigata University, Ikarashi-nihoncho 8050, Nishi-ku, Niigata 950-2181, Japan}

\author[0000-0002-6907-0926]{Kei E. I. Tanaka} 
\affiliation{Department of Earth and Planetary Sciences, Tokyo Institute of Technology, Meguro, Tokyo, 152-8551, Japan}
\affiliation{Center for Astrophysics and Space Astronomy, University of Colorado Boulder, Boulder, CO 80309, USA}
\affiliation{National Astronomical Observatory of Japan, Osawa 2-21-1, Mitaka, Tokyo 181-8588, Japan}

\author[0000-0001-7511-0034]{Yichen Zhang} 
\affiliation{Department of Astronomy, University of Virginia, Charlottesville, VA 22904-4325, USA}
\affiliation{Star and Planet Formation Laboratory, RIKEN Cluster for Pioneering Research, Wako, Saitama 351-0198, Japan}

\author[0000-0002-2026-8157]{Kenji Furuya} 
\affiliation{National Astronomical Observatory of Japan, Osawa 2-21-1, Mitaka, Tokyo 181-8588, Japan}



\begin{abstract}
We report the first detection of hot molecular cores in the Small Magellanic Cloud (SMC), a nearby dwarf galaxy with 0.2 solar metallicity. 
We observed two high-mass young stellar objects in the SMC with the Atacama Large Millimeter/submillimeter Array, and detected emission lines of CO, HCO$^+$, H$^{13}$CO$^+$, SiO, H$_2$CO, CH$_3$OH, SO, and SO$_2$. 
Compact hot-core regions are traced by SO$_2$, whose spatial extent is about $0.1{\rm\:pc}$, and the gas temperature is higher than 100 K based on the rotation diagram analysis. 
In contrast, CH$_3$OH, a classical hot-core tracer, is dominated by extended ($\sim$0.2--0.3${\rm\:pc}$) components in both sources, and the gas temperature is estimated to be 39$^{+8}_{-6}$ K for one source. 
Protostellar outflows are also detected from both sources as high-velocity components of CO. 
The metallicity-scaled abundances of SO$_2$ in hot cores are comparable among the SMC, Large Magellanic Cloud (LMC), and Galactic sources, suggesting that the chemical reactions leading to SO$_2$ formation would be regulated by elemental abundances. 
On the other hand, CH$_3$OH shows a large abundance variation within SMC and LMC hot cores. 
The diversity in the initial condition of star formation (e.g., degree of shielding, local radiation field strength) may lead to the large abundance variation of organic molecules in hot cores. 
This work, in conjunction with previous hot-core studies in the LMC and outer/inner Galaxy, suggests that the formation of a hot core would be a common phenomenon during high-mass star formation across the metallicity range of $0.2$--$1{\rm\:}Z_\odot$. 
High-excitation SO$_2$ lines will be a useful hot-core tracer in the low-metallicity environments of the SMC and LMC.
\end{abstract} 




\section{Introduction} \label{sec_intro} 
Hot molecular cores are compact ($\lesssim$0.1 pc), dense ($\gtrsim$10$^6$ cm$^{-3}$), and hot ($\gtrsim$100 K) protostellar sources, which appear in the early evolutionary stage of massive star formation \citep[e.g.,][]{Kur00}. 
By the observations of Galactic star-forming regions,
a variety of molecular species, including complex organic molecules (COMs), are often detected in hot cores owing to their chemically-rich nature \citep[e.g.,][]{Her09}. 
Obtaining a comprehensive view of chemical compositions of hot cores at various metallicities should help understand the effect of the metallicity on the chemical evolution of protostellar sources. 
Such studies are also important for understanding the physical and chemical processes of the interstellar medium (ISM) during star-/planet-formation in primordial galaxies. 

The characteristic chemistry of a hot core is triggered by radiation from the protostar, which warm up the inner envelope and sublimate ice mantles formed in the earlier colder evolutionary stage. 
Across the ice-sublimation radius, there is a sharp change in molecular abundances between the inner hot-core region ($\gtrsim$100 K) and the outer envelope ($\lesssim$100 K) \citep[e.g.,][]{Cha92,NM04,Gar08b}. 
In lower-metallicity cores, if the grain size distribution is the same, the relative importance of solid-phase chemistry to gas-phase chemistry would be smaller because of the reduced total grain surface area and higher dust temperature. 
Thus, at lower metallicity, we naively expect that hot core regions (i.e., regions where chemistry is dominated by ice sublimation in the vicinity of protostars) would become less distinguished from outer envelopes. 
The validity of such a simplified picture, however, should be examined carefully since many physical and chemical factors contribute to the observed features of hot cores.

With the advent of the Atacama Large Millimeter/submillimeter Array (ALMA), hot cores are now detected in low-metallicity environments of the Large Magellanic Cloud (LMC, $\sim$1/2-1/3 $Z_\odot$) \citep{ST16,ST20,Sew18,Sew22a,Sew22b} and the extreme outer Galaxy ($\sim$1/4 $Z_\odot$) \citep{ST21}. 
Among LMC hot cores, the abundance of CH$_3$OH, one of the simplest COMs and a classical hot-core tracer, shows a large scatter: it is roughly a metallicity-scaled abundance of Galactic counterparts in some sources, while it is significantly depleted beyond the level of metallicity difference in other sources.
In contrast, sulfur-bearing molecules, SO and SO$_2$, are commonly detected in LMC hot cores with similar abundances. 
However, the number of low-metallicity hot core samples is currently still limited, and
observational efforts to extend hot core studies to even lower metallicity environments are highly required.

The Small Magellanic Cloud (SMC) is a nearby star-forming dwarf galaxy located at the distance of 62.1 $\pm$ 1.9 kpc \citep{Gra14}. 
The SMC harbors the lowest metallicity environment \citep[$\sim$1/4-1/10 $Z_\odot$, e.g.,][]{Hun05, Cho18} among the Local-Group star-forming galaxies where we can detect and spatially resolve dense gas tracers. 
It thus provides a valuable opportunity to study chemical processing triggered by star formation in a metal-poor environment.  
However, although the detection of several large molecules such as CH$_3$OH and SO$_2$ are reported for a cold molecular cloud core in the SMC \citep{ST18}, a hot core is yet to be detected due to the lack of a systematic survey. 

In this work, we report the first detection of hot molecular cores in the SMC based on observations with ALMA.

\section{Targets, Observations, and data reduction} \label{sec_tarobsred} 
\subsection{Targets} \label{sec_tar} 
The present targets, S07 and S09 from our our survey catalogue (see \S\ref{sec_obs}) (also known as 2MASS J00540342-7319384 and 2MASS J00445643-7310109, respectively) are classified as high-mass young stellar objects (YSOs) based on their infrared spectral characteristics \citep{vanL10_b}.
Their bolometric luminosities are estimated to be 2.8 $\times$ 10$^4$ $L_{\sun}$ for S07 and 6.1 $\times$ 10$^4$ $L_{\sun}$ for S09, respectively \citep{Oli13}. 
The absorption bands of silicate dust, H$_2$O ice, and CO$_2$ ice are detected in both sources, suggesting that they are still embedded in a dense core \citep{Oli13}. 
The location of the sources within the SMC is shown in Appendix \ref{sec_app_obs}.

\subsection{Observations} \label{sec_obs} 
Observations were made with ALMA in 2019 and 2022 as a Cycle 7 program (2019.1.01770.S, PI: K. Tanaka), which is a part of the ``Magellanic Clouds Outflow and Chemistry Survey (MAGOS)'' project,
in which we observed 10 sources in the SMC and 30 sources in the LMC.
We also use the ALMA archival data, 2019.1.00534.S (PI: S. Zahorecz). 
The sky frequency of 337.81--338.74, 343.59--344.95, 345.01--347.82, 350.44--352.10, 356.15--359.80 GHz, are covered by nine spectral windows in total with a velocity resolution of 0.40--0.43 km s$^{-1}$. 
Details of the observation settings are summarized in Appendix \ref{sec_app_obs}.

\subsection{Data reduction} \label{sec_red} 
Raw data is processed with the \textit{Common Astronomy Software Applications} (CASA) package (version 5.6.1 and 6.2.1). 
The original synthesized beam size is 0$\farcs$37--0$\farcs$43 $\times$ 0$\farcs$30--0$\farcs$33 with the Briggs weighting and a robustness parameter of 0.5. 
For the spectral analysis in this work, all the images are convolved to have a common circular beam size of 0$\farcs$43, which corresponds to 0.13 pc at the distance of the SMC. 
The CASA task tclean is used for imaging and the masking is done using the auto-multithresh algorithm. 
The synthesized images are corrected for the primary beam pattern using the impbcor task in CASA. 
The continuum image is constructed by selecting line-free channels. 
Before the spectral extraction, the continuum emission is subtracted from the spectral data using the uvcontsub task in CASA. 

The spectra and continuum flux are extracted from the 0$\farcs$43 diameter circular region centered at RA = 0$^\mathrm{h}$54$^\mathrm{m}$3$\fs$443 and Dec = -73$\arcdeg$19$\arcmin$38$\farcs$51 (ICRS) for S07 and RA = 0$^\mathrm{h}$44$^\mathrm{m}$56$\fs$421 and Dec = -73$\arcdeg$10$\arcmin$11$\farcs$20 (ICRS) for S09, respectively. 
These positions correspond to the continuum and molecular line peaks of the sources.

\section{Results and analysis} \label{sec_res} 
\subsection{Spectra} \label{sec_spc}  
We identify spectral lines with the aid of the Cologne Database for Molecular Spectroscopy\footnote{https://www.astro.uni-koeln.de/cdms} \citep[CDMS,][]{Mul01,Mul05} and the molecular database of the Jet Propulsion Laboratory\footnote{http://spec.jpl.nasa.gov} \citep[JPL,][]{Pic98}.
Extracted spectra are shown in Appendix \ref{sec_app_fit}.
The detection criterion adopted here is 3$\sigma$ significance level and the velocity coincidence with the systemic velocities of S07 (163.1 km s$^{-1}$) and S09 (132.0 km s$^{-1}$), which are measured using the SO($N_J$ = 8$_{9}$--7$_{8}$) line. 
For weak lines, two or three channels are binned to increase the sensitivity. 

Molecular emission lines of CO, HCO$^{+}$,  H$^{13}$CO$^{+}$, SiO, H$_2$CO, CH$_3$OH, SO, and SO$_2$ are detected in both S07 and S09. 
For SO$_2$ and CH$_3$OH, multiple high excitation lines with different upper state energies are detected. 
Line parameters are measured by fitting a Gaussian. 
We estimate the peak brightness temperature, the FWHM, the LSR velocity, and the integrated intensity for each line based on the fitting as summarized in Appendix \ref{sec_app_line}.

\subsection{Images} \label{sec_img} 
\begin{figure*}[tp!]
\begin{center}
\includegraphics[width=17.5cm]{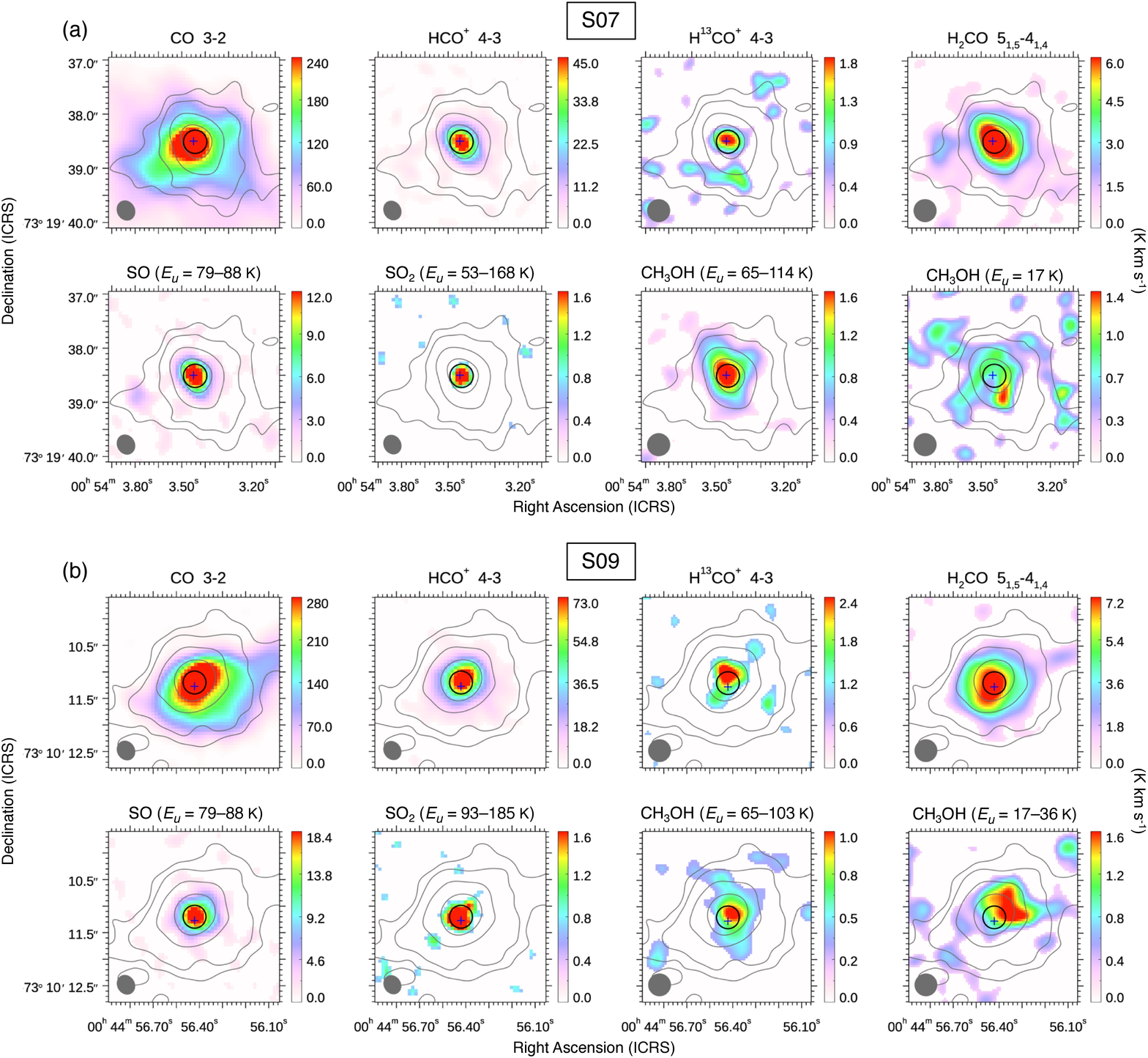}
\caption{
Integrated intensity distributions of molecular emission lines. 
Gray contours represent the 850 $\mu$m continuum distribution and the contour levels are 3$\sigma$, 6$\sigma$, 15$\sigma$, 40$\sigma$ of the rms noise (0.03-0.04 mJy/beam). 
Low signal-to-noise ratio regions (S/N $<$1.5) are masked. 
The spectra discussed in the text are extracted from the region indicated by the black open circle. 
The blue cross represents the infrared peak position. 
The synthesized beam size is shown by the gray filled circle in each panel. 
North is up, and east is to the left. 
}
\label{images}
\end{center}
\end{figure*}

Figure \ref{images} shows synthesized images of continuum and molecular emission lines. 
The images are constructed by integrating spectral data in the velocity range where the emission is detected. 
For SO$_2$ and CH$_3$OH, several emission lines with similar upper state energies are stacked to reduce the noise level. 
Almost molecular emission have their intensity peak near the submillimeter continuum center, which also corresponds to the infrared peak of the sources. 
SiO emission is not shown in the figure since the line is weak, but it also shows a compact distribution at the continuum peak. 
We estimate the size of emission regions by fitting a two-dimensional Gaussian (see Table \ref{tab_N}). 
We also estimated the deconvolved source sizes as 
$\sqrt{\mathrm{FWHM_{maj}} \times \mathrm{FWHM_{min}} - \Theta_{\mathrm{beam}}^2 }$, 
where $\mathrm{FWHM_{maj/min}}$ is the major/minor FWHM size as estimated from the two-dimensional Gaussian fit, 
and $\Theta_{\mathrm{beam}}^2$ is the geometric mean of the beam major and minor axes. 
Note that only upper limits are estimated for the emission whose size is comparable with or smaller than the beam size. 
Details of the continuum and molecular line distributions are discussed in Section \ref{sec_disc_star}.

\begin{deluxetable*}{ l c c c c c c }
\tablecaption{Estimated rotation temperatures, column densities, and source sizes \label{tab_N}}
\tabletypesize{\small} 
\tablehead{
\colhead{Object}                       & \colhead{$T$$_{rot}$}   &       \colhead{$N$(X)}            &  \colhead{$N$(X)/$N_{\mathrm{H_2}}$}   &  Size\tablenotemark{a}     & Size$_{\mathrm{deconv}}$\tablenotemark{b}  & Note  \\
\colhead{/ Molecule}                & \colhead{(K)}                 &        \colhead{(cm$^{-2}$)}     & \colhead{}                                                 & \colhead{($\arcsec$/pc)}  & \colhead{($\arcsec$/pc)}      & \colhead{}  
}
\startdata 
S07 & & & & & & \\
 H$_2$                                         &   \nodata                     &   (1.5 $\pm$ 0.5) $\times$ 10$^{23}$               & \nodata                                                & 0.65/0.20\tablenotemark{c}    & 0.55/0.17\tablenotemark{c}           &  (1)   \\
 SO$_2$                                      &   171$^{+65}_{-37}$   &     (3.0 $^{+0.7}_{-0.5}$) $\times$ 10$^{14}$   & (2.0 $\pm$ 0.8) $\times$ 10$^{-9}$      & 0.32/0.10                                   & $\lesssim$0.17/0.05\tablenotemark{d}  &  (2)  \\
 CH$_3$OH                                 &   39$^{+8}_{-6}$        &    (2.8 $^{+1.0}_{-0.7}$) $\times$ 10$^{14}$     &  (1.8 $\pm$ 0.9) $\times$ 10$^{-9}$     & 0.82/0.25\tablenotemark{e}    & 0.70/0.21\tablenotemark{e}         &  (2) \\
 SO                                              &   30--60                       &    (3.5 $\pm$ 1.4) $\times$ 10$^{14}$              &  (2.3 $\pm$ 1.0) $\times$ 10$^{-9}$    & 0.48/0.15                                   & 0.32/0.10                                   &  (3) \\
 SiO                                             &   30--60                       &    (3.1 $\pm$ 0.9) $\times$ 10$^{12}$               & (2.1 $\pm$ 0.9) $\times$ 10$^{-11}$   & \nodata                                     & \nodata                                      &  (3) \\ 
 H$_2$CO                                   &   30--60                        &    (3.6 $\pm$ 0.1) $\times$ 10$^{13}$              &  (2.4 $\pm$ 0.8) $\times$ 10$^{-10}$   & 0.89/0.27                                & 0.78/0.24                                    &   (3) \\
 HCO$^+$                                   &   30--60                        &    (2.0 $\pm$ 0.1) $\times$ 10$^{13}$              &  (1.3 $\pm$ 0.4) $\times$ 10$^{-10}$   & 0.55/0.17                                & 0.26/0.08                                    &  (3) \\
 HCO$^+$ ($^{13}$C)                 &   \nodata                      &    (5.5 $\pm$ 1.7) $\times$ 10$^{13}$              &  (3.7 $\pm$ 1.1) $\times$ 10$^{-10}$   & \nodata                                    & \nodata                                      &  (4) \\
 H$^{13}$CO$^+$                       &   30--60                        &    (1.1 $\pm$ 0.1) $\times$ 10$^{12}$              &  (7.3 $\pm$ 2.5) $\times$ 10$^{-12}$   & 0.44/0.13                                & $\lesssim$0.22/0.07\tablenotemark{d}  &  (3) \\
 CO                                              &   30--60                        & $>$2 $\times$ 10$^{17}$                                 &  $>$1 $\times$ 10$^{-6}$                     & 1.27/0.38                                & 1.22/0.37                                    &  (5)  \\
\tableline
S09 & & & & & & \\
 H$_2$                                         &   \nodata                     &   (2.5 $\pm$ 1.0) $\times$ 10$^{23}$               & \nodata                                                 & 0.68/0.21\tablenotemark{c}  & 0.58/0.18\tablenotemark{c}      &  (1)   \\
 SO$_2$ ($E_{u} >$80 K)           &   146$^{+68}_{-35}$    &     (1.9 $^{+0.7}_{-0.5}$) $\times$ 10$^{14}$   & (7.2 $\pm$ 5.9) $\times$ 10$^{-10}$    & 0.46/0.14                                 & 0.30/0.09                               &  (6)  \\
 SO$_2$ ($E_{u} <$60 K)           &    17$^{+6}_{-4}$         &     (1.8 $^{+1.7}_{-0.9}$) $\times$ 10$^{14}$   & (7.6 $\pm$ 3.9) $\times$ 10$^{-10}$    & \nodata                                    & \nodata                                 &  (6)  \\
 CH$_3$OH                                 & 30--60                        &    (1.6 $\pm$ 0.5) $\times$ 10$^{14}$             &  (6.4 $\pm$ 3.2) $\times$ 10$^{-10}$    & 0.93/0.28\tablenotemark{e}   & 0.82/0.25\tablenotemark{e}    &  (7) \\
 SO                                              &   30--60                       &    (3.5 $\pm$ 1.1) $\times$ 10$^{14}$              &  (1.4 $\pm$ 0.7) $\times$ 10$^{-9}$     & 0.58/0.18                               & 0.46/0.14                                &  (3) \\
 SiO                                             &   30--60                       &    $<$2 $\times$ 10$^{12}$                              & $<$8 $\times$ 10$^{-12}$                     & \nodata                                  & \nodata                                   &  (3) \\ 
 H$_2$CO                                   &   30--60                        &    (3.4 $\pm$ 0.1) $\times$ 10$^{13}$              &  (1.4 $\pm$ 0.5) $\times$ 10$^{-10}$   & 1.06/0.32                              & 0.97/0.30                                 &  (3) \\
 HCO$^+$                                   &   30--60                        &    (2.9 $\pm$ 0.1) $\times$ 10$^{13}$              &  (1.2 $\pm$ 0.5) $\times$ 10$^{-10}$   & 0.69/0.21                               & 0.60/0.18                               &  (3) \\
 HCO$^+$ ($^{13}$C)                 &   \nodata                      &    (5.3 $\pm$ 1.6) $\times$ 10$^{13}$              &  (2.2 $\pm$ 0.7) $\times$ 10$^{-10}$   & \nodata                                  & \nodata                                   &  (4) \\
 H$^{13}$CO$^+$                       &   30--60                        &    (1.1 $\pm$ 0.1) $\times$ 10$^{12}$              &  (4.4 $\pm$ 1.9) $\times$ 10$^{-12}$   & 0.57/0.17                              & 0.38/0.11                               &  (3) \\
 CO                                              &   30--60                        & $>$2 $\times$ 10$^{17}$                                 &  $>$8 $\times$ 10$^{-7}$                    & 1.25/0.38                               & 1.20//0.36                               &  (5)  \\
\enddata
\tablecomments{
Uncertainties and upper limits are of 2$\sigma$ level and do not include systematic errors due to adopted spectroscopic constants. \\
(1) Based on dust continuum. $T_{d}$ is varied from 50 to 100 K to estimate the range of possible $N_{\mathrm{H_2}}$. 
(2) Based on the rotation diagram. 
(3) $T$$_{rot}$ is varied from 30 to 60 K to estimate the range of possible $N$(X). 
(4) Estimated from H$^{13}$CO$^+$ assuming $^{12}$C/$^{13}$C = 50 \citep{Hei99}. An empirical 30 $\%$ uncertainty is assumed. 
(5) Only lower limits since the line is optically thick. 
(6) Based on the rotation diagram analysis with two temperature components ($E_{u} >$80 K and $E_{u} <$60 K). 
(7) Estimated using CH$_3$OH(7$_{-1}$ E--6$_{-1}$ E). 
}
\tablenotetext{a}{Source sizes estimated from a two-dimensional Gaussian fit to the emission peak (geometric mean of major and minor FWHM sizes). }
\tablenotetext{b}{Deconvolved source size. Major and minor beam sizes range from 0$\farcs$43-0$\farcs$37 and 0$\farcs$43-0$\farcs$32. See Section \ref{sec_img} for details. }
\tablenotetext{c}{Size of dust continuum emission. }
\tablenotetext{d}{We set a minimum deconvolved size to be half the beam size to avoid unrealistically small source sizes. }
\tablenotetext{e}{Size of high-$E_{u}$ lines. Low-$E_{u}$ lines are more extended. }
\end{deluxetable*}

\subsection{Column densities and gas temperatures} \label{sec_rd}
\begin{figure*}[tp!]
\begin{center}
\includegraphics[width=18cm]{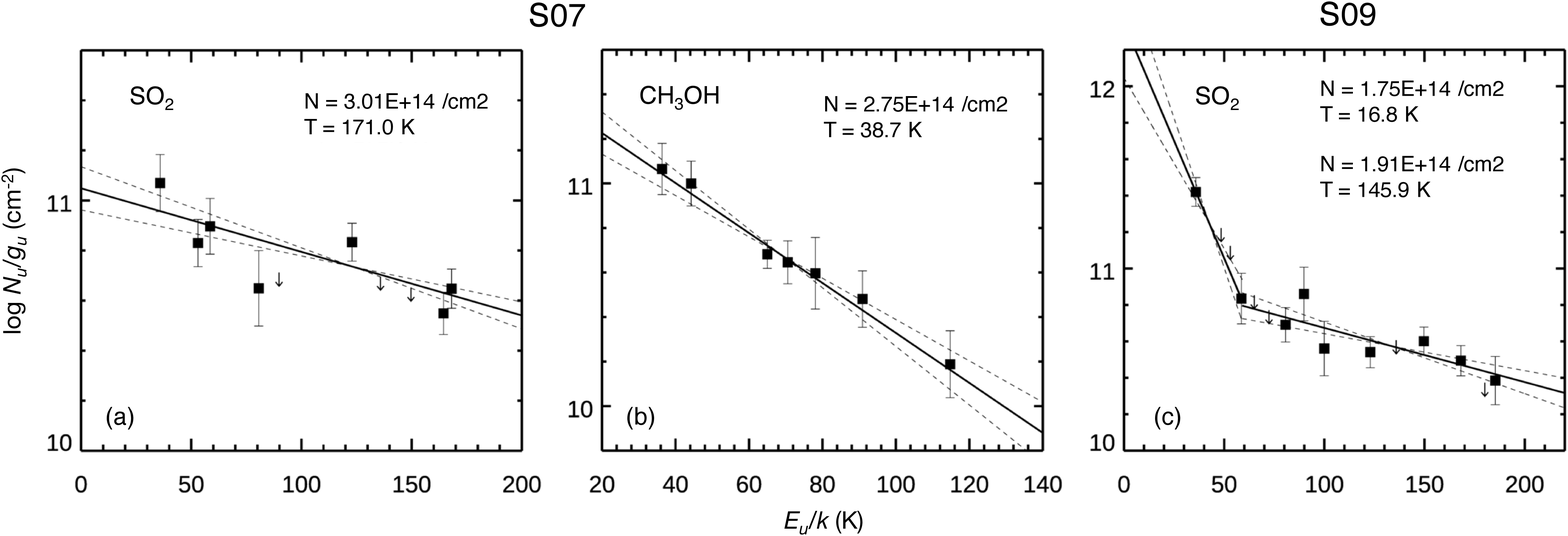}
\caption{
Results of rotation diagram analyses. 
Upper limit points are indicated by the downward arrows. 
The solid lines represent the fitted straight line, while the dashed lines indicate the acceptable fits within 2$\sigma$ level. 
Derived column densities and rotation temperatures are shown in each panel. 
See Section \ref{sec_rd} for details. 
}
\label{rd1}
\end{center}
\end{figure*}

We use the rotation diagram method to estimate column densities and temperatures of SO$_2$ in both sources and CH$_3$OH in S07 (Figure \ref{rd1}), while the number of detected CH$_3$OH lines in S09 is not sufficient for this purpose.
The details of the formulae are described in Appendix \ref{sec_app_rd}.
For SO$_2$ in S09, a straight-line fit is separated into low- and high-temperature regimes ($E_{u}$ $<$60 K and $E_{u}$ $>$60 K), because different temperature components are clearly seen in the diagram.
The high-temperature components of SO$_2$ with $T_{\rm rot}>100{\rm\:K}$ are detected in both sources.
Derived parameters are summarized in Table \ref{tab_N}.

For other molecules, the rotation temperatures are assumed for the column density estimation.
We have applied low rotation-temperatures from 30 to 60 K, because the analyses for hot cores in the LMC and the outer Galaxy report $T_{\rm rot}\sim$40--50 K for relatively small molecules with extended spatial distributions, such as SO and CS \citep{ST20,ST21}. 
The same temperature range is applied for CH$_3$OH in S09, which also shows an extended distribution.
If the rotation temperature of CH$_3$OH in S09 is assumed to be the same as that of the high-temperature component of SO$_2$, then we obtain a twice higher CH$_3$OH column density. 

\subsection{H$_2$ column density and abundances } \label{sec_h2} 
The column density of molecular hydrogen ($N_{\mathrm{H_2}}$) is estimated from the dust continuum data based on the standard treatment of optically thin emission (see Appendix \ref{sec_app_h2} for details). 
The assumption of the dust temperature $T_{d}$ may cause considerable uncertainty in the derivation of $N_{\mathrm{H_2}}$. 
A detailed analysis of effective dust temperature in the line of sight toward an LMC hot core suggests that $T_{d}$ is somewhat lower than the temperature of a hot core region traced by SO$_2$ or CH$_3$OH due to the temperature gradient in a protostellar envelope \citep{ST20}. 
In this work, we have applied $T_{d}$ from 50 to 100 K to estimate the possible range of the $N_{\mathrm{H_2}}$ values. 
The 850 $\mu$m continuum brightness is measured to be 1.52 mJy/beam for S07 and 2.62 mJy/beam for S09 (with a beam size of 0$\farcs$37 $\times$ 0$\farcs$32), and with the above assumption, 
we obtain $N_{\mathrm{H_2}}$ = (1.5 $\pm$ 0.5) $\times$ 10$^{23}$ cm$^{-2}$ for S07 and (2.5 $\pm$ 0.8) $\times$ 10$^{23}$ cm$^{-2}$ for S09, respectively. 
Fractional molecular abundances with respect to H$_2$ are calculated using those $N_{\mathrm{H_2}}$ and summarized in Table \ref{tab_N}. 
The gas number density is estimated to be $n_{\mathrm{H_2}}$ = 6 $\times$ 10$^5$ cm$^{-3}$ for S07 and 9 $\times$ 10$^5$ cm$^{-3}$ for S09, assuming the source diameter of 0.13 pc and the uniform spherical distribution of gas around a protostar.

\section{Discussion} \label{sec_disc} 
\subsection{Hot molecular cores associated with S07 and S09} \label{sec_disc_star} 
S07 and S09 show the following characteristics; 
(i) the compact distribution of dense gas tracers ($\sim$0.1 pc, Section \ref{sec_img}), 
(ii) the high gas temperature ($\geq$100 K, Section \ref{sec_rd}), 
(iii) the high density ($\sim$10$^6$ cm$^{-3}$, Section \ref{sec_h2}), 
(iv) the association with a high-mass protostar (3-6 $\times$ 10$^4$ $L_{\sun}$), 
and 
(v) the association with protostellar outflows (see Appendix \ref{sec_app_outflow}). 
These natures suggest that the sources are still in the early evolutionary stage of high-mass star formation, harboring a hot molecular core in the vicinity of the central protostar.

\subsection{SO$_2$ as a tracer of low-metallicity hot cores} \label{sec_disc_so2} 
In both sources, a compact high-temperature region around the protostar is traced by SO$_2$ gas. 
The deconvolved source size of SO$_2$ is $\lesssim$0.05 pc for S07 and 0.09 pc for S09 (Table. \ref{tab_N}), which are comparable with or smaller than those of known LMC hot cores \citep{ST16b,Sew18,ST20,Sew22b}. 
With the LTE assumption, the compact high-temperature regions in S07 and S09 are hot enough to sublimate the ice mantle,
and the SO$_2$ gas likely arises mainly from the hot cores.

In contrast, the CH$_3$OH emission show larger deconvolved source sizes (0.21 pc for S07 and 0.25 pc for S09) compared to that of SO$_2$, and associated with extended emission components.
Although several high-$E_{u}$ lines ($E_{u}$ $>$ 60 K) in S07 are detected, the rotation temperature is much lower than that of SO$_2$.
Low-excitation CH$_3$OH lines ($E_{u}$ = 17--36 K) are even extended, and the emission peak does not correspond to the hot core position. 

The clearly different spatial distributions and rotation temperatures between SO$_2$ and CH$_3$OH are in stark contrast to the characteristics of known LMC hot cores, as their high-excitation SO$_2$ and CH$_3$OH lines are both compact ($\sim$0.1 pc) and hot ($T$$_{rot}$ $>$100 K) \citep[e.g.,][]{ST20,Sew22b}.
In the present SMC hot cores, SO$_2$ and CH$_3$OH may have different physical origins. 

Both SO$_2$ and CH$_3$OH are believed to be tracer molecules of hot cores in the Galactic and LMC star-forming regions. 
In hot cores, SO$_2$ could be formed by high-temperature gas-phase reactions of H$_2$S, HS, and OCS, which are sublimated from sulfur-bearing ices \citep[e.g.,][]{Vid18}.
If solid H$_2$S is a dominant sulfur reservoir in a prestellar stage, the expected major pathway to form SO$_2$ in a hot core is
H$_2$S $\xrightarrow{\mathrm{H}}$ SH $\xrightarrow{\mathrm{H}}$ S $\xrightarrow{\mathrm{OH/O_2}}$ SO $\xrightarrow{\mathrm{OH}}$ SO$_2$, 
and the co-evolution of SO and SO$_2$ abundances is predicted by astrochemical simulations \citep[e.g.,][]{Cha97, NM04}. 
Alternatively, SO$_2$ can be produced at a prestellar stage by cosmic-ray-induced radiation chemistry and bulk reactions of reactive species in ice mantles \citep{Shin20}, and subsequently released into gas-phase by sublimation in a hot core. 

\begin{figure}[tpb!]
\begin{center}
\includegraphics[width=7.5cm]{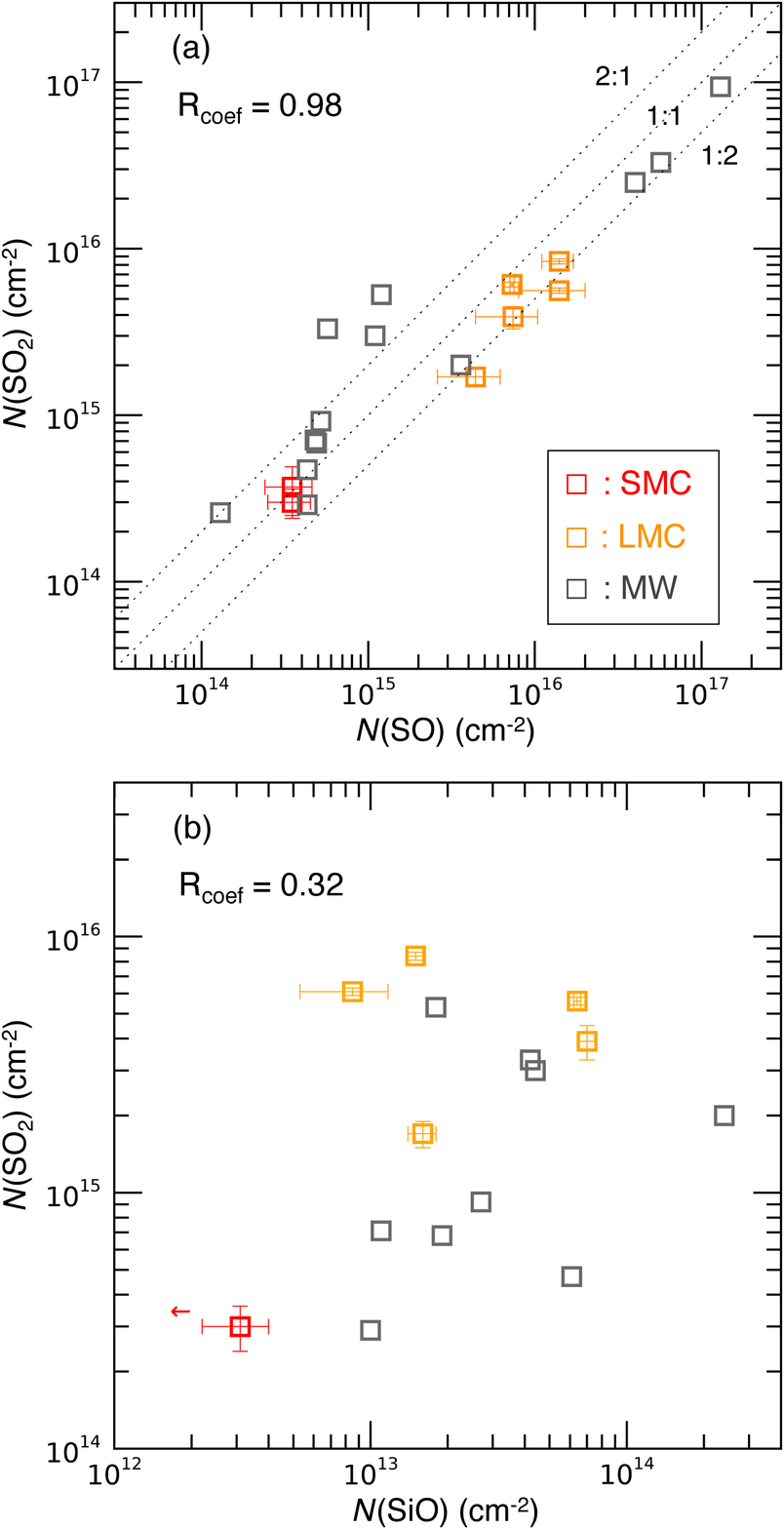}
\caption{
SO$_2$ column densities plotted against those of SO (a) and SiO (b) for SMC (red), LMC (orange), and Galactic (gray) hot cores. 
The dotted lines in panel (a) represent a ratio of 2:1, 1:1, and 1:2. 
The correlation coefficient (R$_\mathrm{coef}$) is indicated. 
}
\label{so2}
\end{center}
\end{figure}

Figure \ref{so2} compares the SO$_2$ column densities with those of SO and SiO for SMC, LMC, and Galactic hot cores (see Section \ref{sec_disc_molab} for the references of the plotted data). 
The SO$_2$ column densities show a tight correlation with SO (correlation coefficient: 0.98), while they do not correlate with the strong-shock tracer SiO (correlation coefficient: 0.32).
This would suggest that the formation of SO$_2$ in the current sample is mainly induced by the sublimation of sulfur-bearing ices, rather than sputtering by strong shocks. 

CH$_3$OH is reported to be efficiently formed by the successive hydrogenation of CO on grain surfaces \citep[e.g.,][]{Wat02}. 
The reaction typically proceeds at low temperatures ($\lesssim$ 20 K), and thus abundant solid CH$_3$OH can be formed in cold prestellar/protostellar stages \citep[e.g.,][]{Boo11,Whi11}. 
At the hot-core stage, the radiation from the protostar can trigger the thermal desorption of ice mantles, and if the abundant reservoir of solid CH$_3$OH is present, the abundance of gaseous CH$_3$OH will be significantly enhanced. 

In the case of S07 and S09, such a high-temperature region is traced by hot SO$_2$ gas with a compact distribution centered at the continuum peak. 
A relatively low temperature and extended spatial distribution of CH$_3$OH suggest that the contribution from a non-thermal desorption is not negligible for its production in both sources.  
Considering the presence of protostellar outflows (Appendix \ref{sec_app_outflow}), sputtering of ice mantles by shock \citep[e.g.,][]{Aot15} or photodesorption by UV photons in an outflow cavity wall of the central protostar may be responsible for the desorption mechanism. 
Note that the latter mechanism is still under debate, since laboratory experiments have reported that solid CH$_3$OH readily dissociates upon photodesorption by the UV irradiation \citep{Ber16,Cru16}.

\begin{figure*}[tpb!]
\begin{center}
\includegraphics[width=12.0cm]{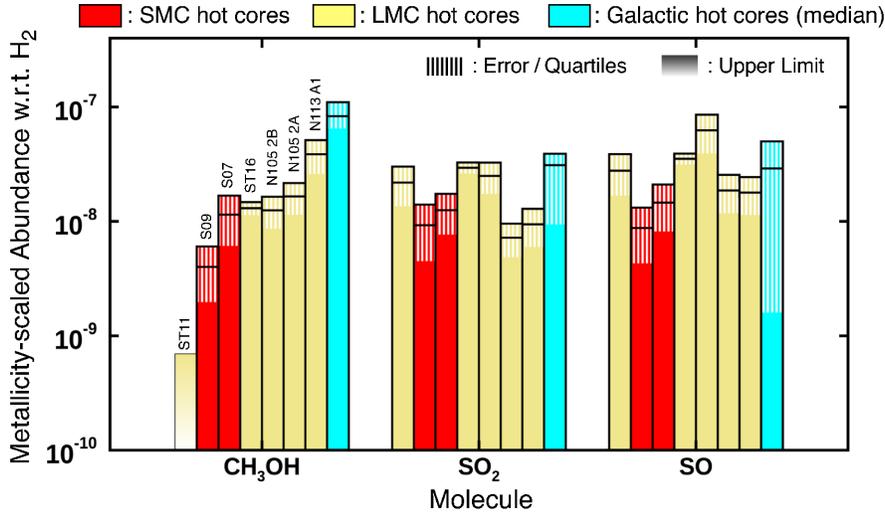}
\caption{
Comparison of metallicity-scaled molecular abundances of hot cores between SMC (red, this work), LMC (yellow), and Galactic sources (cyan). 
Abundances of SMC and LMC hot cores are multiplied by the corresponding metallicity factors (see Section \ref{sec_disc_molab}). 
The area with thin vertical lines indicate the error bar for SMC/LMC sources, while those of Galactic sources indicate the lower and upper quartiles of the abundance distribution. 
The bar with a color gradient indicate an upper limit. 
See Section \ref{sec_disc_molab} for details. 
}
\label{histo}
\end{center}
\end{figure*}

\subsection{Molecular abundances} \label{sec_disc_molab} 
From the point of view of molecular abundances, SO$_2$ and CH$_3$OH show different source-to-source variations within LMC/SMC hot cores. 
Figure \ref{histo} shows a comparison of metallicity-scaled molecular abundances for SMC, LMC, and Galactic hot cores. 
The data of LMC hot cores are adopted from \citet{ST16,ST20,Sew22b}, where the $N_{\mathrm{H_2}}$ of ST11, N105 2B/2A, and N113 A1 are re-estimated using the same dust opacity data as in this work and dust temperature of $T_{d}$ = 60 K. 
Abundances for Galactic sources are collected from the literatures for a sample of 11--17 hot cores \citep{Sut95, Mac96, Hel97, Hat98b, vdT00, Qin10, Zer12, Xu13, Ger14}. 
In the figure, the lower and upper quartiles of the abundance distributions are plotted with the median value. 
HCO$^+$ abundances are calculated from H$^{13}$CO$^+$ assuming $^{12}$C/$^{13}$C = 49, 50, and 77 for SMC, LMC, and Galactic sources \citep{Hei99, Wan09, Wil94}. 

In Figure \ref{histo}, to correct for the metallicity difference, the molecular abundances of SMC and LMC hot cores are scaled in accordance with their metallicities. 
We have estimated the local metallicities of the individual hot cores based on the metallicity maps of the SMC and LMC reported in \citet{Cho16,Cho18}. 
The metallicity for the two SMC hot cores is measured to be [Fe/H] $=$ $-$0.8 dex (0.2 $Z_\odot$), while that for the LMC hot cores is typically  [Fe/H] $=$ $-$0.4 dex (0.4 $Z_\odot$). 
We assume a solar metallicity for Galactic hot cores. 

As seen in the figure, the CH$_3$OH abundance shows large source-to-source variation even after corrected for the metallicity. 
In contrast, the metallicity-scaled abundance of SO$_2$ is relatively constant among SMC, LMC, and Galactic sources. 
The reason for the large abundance diversity of CH$_3$OH gas in LMC/SMC hot cores is still under debate. 
Astrochemical simulations for the chemical evolution of LMC/SMC hot cores suggest that dust temperature at the initial ice-forming stage has a significant effect on the CH$_3$OH gas abundance in the subsequent hot core stage, because the formation of solid CH$_3$OH is sensitive to dust temperature and it is inhibited on warmer grain surfaces \citep{Ach18,ST20}. 
Therefore, the diversity in the initial condition of star formation (e.g., degree of shielding, local radiation field strength) may lead to the large abundance variation of CH$_3$OH. 
Infrared ice observations argue that the inhibition of solid CH$_3$OH in the ice-forming stage is more likely to occur in the LMC condition due to the lower dust abundance and the stronger interstellar radiation field \citep{ST16}.

On the other hand, SO$_2$ would not be inherited from the prestellar stage to the hot core stage, because it is mainly synthesized by high-temperature gas-phase chemistry. 
Thus, except for the elemental abundance effect, the local interstellar condition would not largely affect the SO$_2$ abundance in hot cores, which may result in a constant metallicity-scaled abundance of SO$_2$ across SMC, LMC, and Galactic sources. 
Such behavior of SO$_2$ is predicted by the astrochemical simulations dedicated to LMC hot cores \citep{ST20}, but now the present study suggests that it is also applicable to the SMC hot cores.

\section{Summary} \label{sec_sum} 
We report the detection of two hot molecular cores in the SMC based on 0.1 pc-scale submillimeter observations toward high-mass YSOs with ALMA. 
Emission lines of CO, HCO$^+$, H$^{13}$CO$^+$, SiO, H$_2$CO, CH$_3$OH, SO, and SO$_2$ are detected from the sources. 
The associated protostellar outflows are also observed. 
The compact hot-core regions are traced by SO$_2$ gas, whose spatial extent is about 0.1 pc, and the gas temperature is higher than 100 K based on the excitation analysis. 
In contrast, CH$_3$OH, a classical hot-core tracer, is dominated by extended ($\sim$0.2--0.3${\rm\:pc}$) components, and low gas temperature ($<$40 K) is inferred for one source. 
The metallicity-scaled abundances of SO$_2$ in hot cores are comparable between SMC, LMC, and Galactic sources, suggesting that the chemical reactions leading to the formation of SO$_2$ would be regulated by the sulfur abundances. 
On the other hand, CH$_3$OH shows a large abundance variation within SMC and LMC hot cores. 
This difference would be due to their different chemical origin. 
High-excitation SO$_2$ lines will be a useful tracer of hot cores in the low-metallicity environments of the SMC and LMC. 
In conjunction with previous hot core studies in the LMC and outer/inner Galaxy, the formation of a hot core would be a common phenomenon during star formation across the metallicity range of 0.2--1 $Z_\odot$.

\vspace{8pt}
{\small 
This paper makes use of the following ALMA data: ADS/JAO.ALMA$\#$2019.1.01770.S and 2019.1.00534.S. 
ALMA is a partnership of ESO (representing its member states), NSF (USA) and NINS (Japan), together with NRC (Canada), MOST and ASIAA (Taiwan), and KASI (Republic of Korea), in cooperation with the Republic of Chile. 
The Joint ALMA Observatory is operated by ESO, AUI/NRAO and NAOJ. 
This work has made extensive use of the Cologne Database for Molecular Spectroscopy and the molecular database of the Jet Propulsion Laboratory. 
This work was supported by JSPS KAKENHI grant Nos. JP19K14760, JP20H05845C, JP20H05847, JP21H00037, JP21H00058, and JP21H01145. 
T. S. was supported by Leading Initiative for Excellent Young Researchers, MEXT, Japan.
Finally, we would like to thank an anonymous referee for insightful comments, which substantially improved this paper.  
}

\software{CASA \citep{McM07})}




\appendix

\restartappendixnumbering

\section{Observation summary} \label{sec_app_obs} 
\begin{figure}[ptb] 
\includegraphics[width=8.0cm]{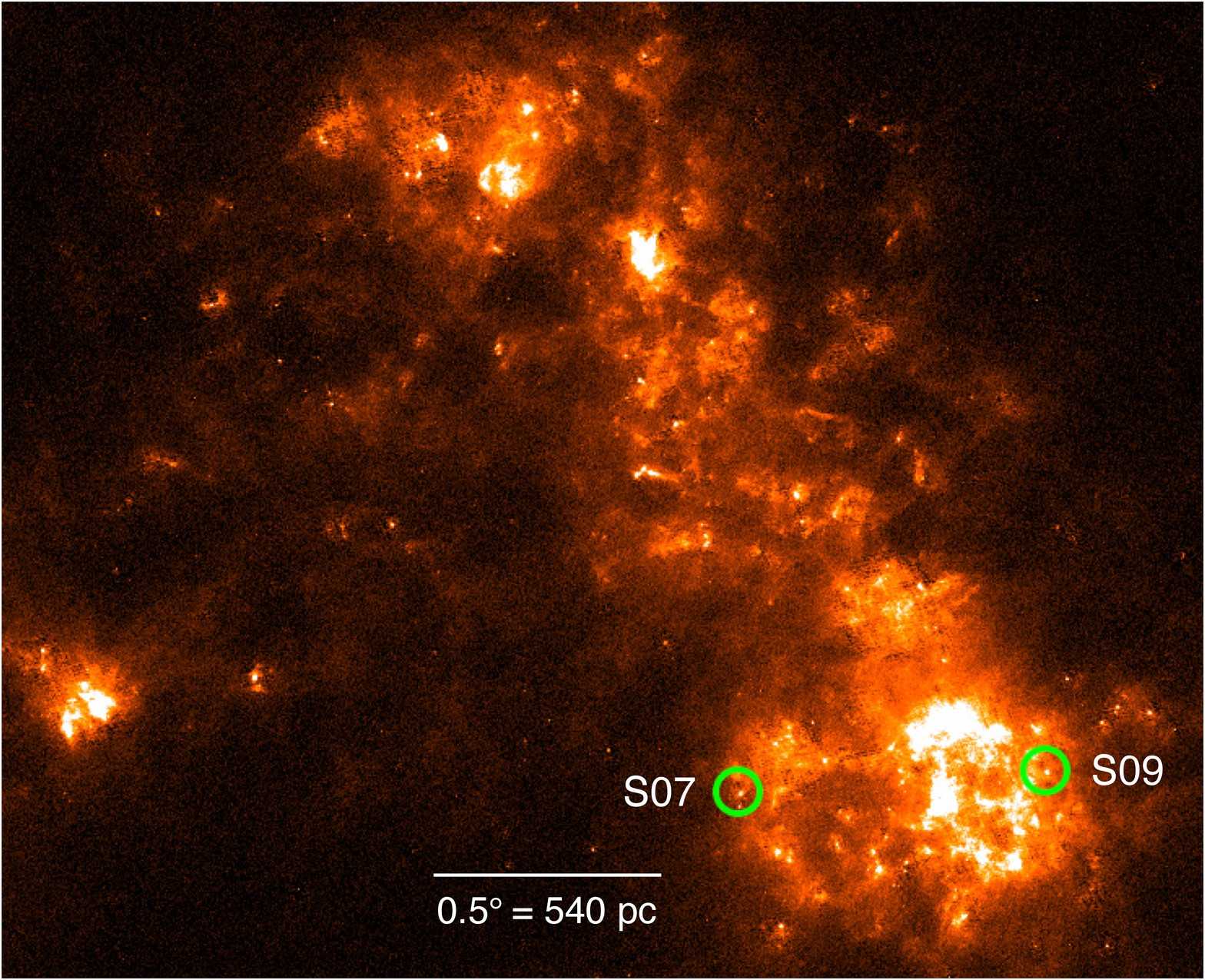} 
\caption{
Location of the present hot cores, S07 and S09, within the SMC (the green solid circle). 
The background is a Herschel/PACS 160 $\mu$m image obtained by the HERITAGE program \citep{Mei13}. 
North is up, and east is to the left. 
}
\label{SMC_whole}
\end{figure}

Table \ref{tab_Obs} summarizes the details of the present ALMA Band 7 observations. 
The location of the present hot cores in the SMC is shown in Figure \ref{SMC_whole}. 

\begin{deluxetable*}{ l c c c c c c c c c}
\tablecaption{Observation summary \label{tab_Obs}} 
\tablewidth{0pt} 
\tabletypesize{\footnotesize} 
\tablehead{
\colhead{Object / }   & \colhead{Observation} &  \colhead{On-source} & \colhead{Mean}                             & \colhead{Number}   &  \multicolumn{2}{c}{Baseline\tablenotemark{b}}     &  \colhead{}                                                  &  \colhead{}                                         &  \colhead{Channel}  \\
\cline{6-7}  
\colhead{Proposal ID}   & \colhead{Date}             &  \colhead{Time}         &  \colhead{PWV\tablenotemark{a}} & \colhead{of}              &  \colhead{L5} & \colhead{L80} & \colhead{Beam size\tablenotemark{c}}        & \colhead{MRS\tablenotemark{d}}     &  \colhead{Spacing}  \\
\colhead{}   & \colhead{}                    &  \colhead{(min)}         &  \colhead{(mm)}                             & \colhead{Antennas} & \colhead{(m)} & \colhead{(m)}   &  \colhead{($\arcsec$ $\times$ $\arcsec$)} &  \colhead{($\arcsec$)}                       &  \colhead{(MHz)}                }
\startdata 
S07       &                                        &                                       &                                                      &                                     &                      &                          &                                                                    &                                         &                \\
2019.1.00534.S       &  2019 Oct 17 -- 2019 Oct 21 &  35.3                           &  0.5--0.9                                         &  42-45                       &  36.0               &  278.1              &  0.41 $\times$ 0.30                                     &   4.9                              &  0.49                \\
                                &                  &                                    &                                                       &                                  &                        &                         &                                                                     &                                      &      \\
2019.1.00534.S       &  2019 Oct 6 -- 2019 Oct 7   &  22.7                           &  0.5                                                 &  39-44                       &  44.5               &  351.2            &  0.39 $\times$ 0.32                                     &   4.0                              &  0.49                \\
                                &                  &                                    &                                                         &                                  &                        &                         &                                                                     &                                      &      \\
2019.1.01770.S       &  2019 Oct 21 -- 2019 Oct 23, &  24.2                           &  0.5--1.1                                         &  44-48                       &  35.2               &  270.7              &  0.36 $\times$ 0.32                                     &   5.0                              &  0.49                \\
                      &  2022 Jan 6, 2022 May 17              &                                    &                                                       &                                  &                        &                         &                                                                     &                                      &      \\
\tableline
S09       &                                       &                                       &                                                      &                                     &                      &                          &                                                                    &                                         &                \\
2019.1.00534.S       &  2019 Oct 17 --            &  36.3                           &  0.5--0.9                                         &  42-45                       &  36.1               &  277.9              &  0.41 $\times$ 0.30                                     &   4.8                              &  0.49                \\
                                &  2019 Oct 21               &                                    &                                                       &                                  &                        &                         &                                                                     &                                      &      \\
2019.1.00534.S       &  2019 Oct 6 --             &  22.7                           &  0.5                                                 &  39-44                       &  44.8               &  351.3            &  0.39 $\times$ 0.32                                     &   4.0                              &  0.49                \\
                                &  2019 Oct 7                &                                    &                                                         &                                  &                        &                         &                                                                     &                                      &      \\
2019.1.01770.S       & 2019 Oct 21 -- 2019 Oct 23, &  24.2                           &  0.5--1.1                                         &  44-48                       &  35.4               &  271.7              &  0.36 $\times$ 0.32                                     &   5.0                              &  0.49                \\
                      &  2022 Jan 6, 2022 May 17  &                                    &                                                       &                                  &                        &                         &                                                                     &                                      &      \\
\enddata
\tablenotetext{a}{Precipitable water vapor.}
\tablenotetext{b}{L5/L80 indicate the length that includes the 5th/80th percentile of all projected baselines. }
\tablenotetext{c}{The average beam size of aggregated continuum. 
Note that we use a common circular restoring beam size of 0$\farcs$43 for the spectral analysis. }
\tablenotetext{d}{Maximum Recoverable Scale.}
\end{deluxetable*}

\section{Fitted spectra and measured line parameters} \label{sec_app_fitting}
\subsection{Fitted spectra} \label{sec_app_fit}
\begin{figure*}[tp] 
\begin{center} 
\includegraphics[width=17.0cm]{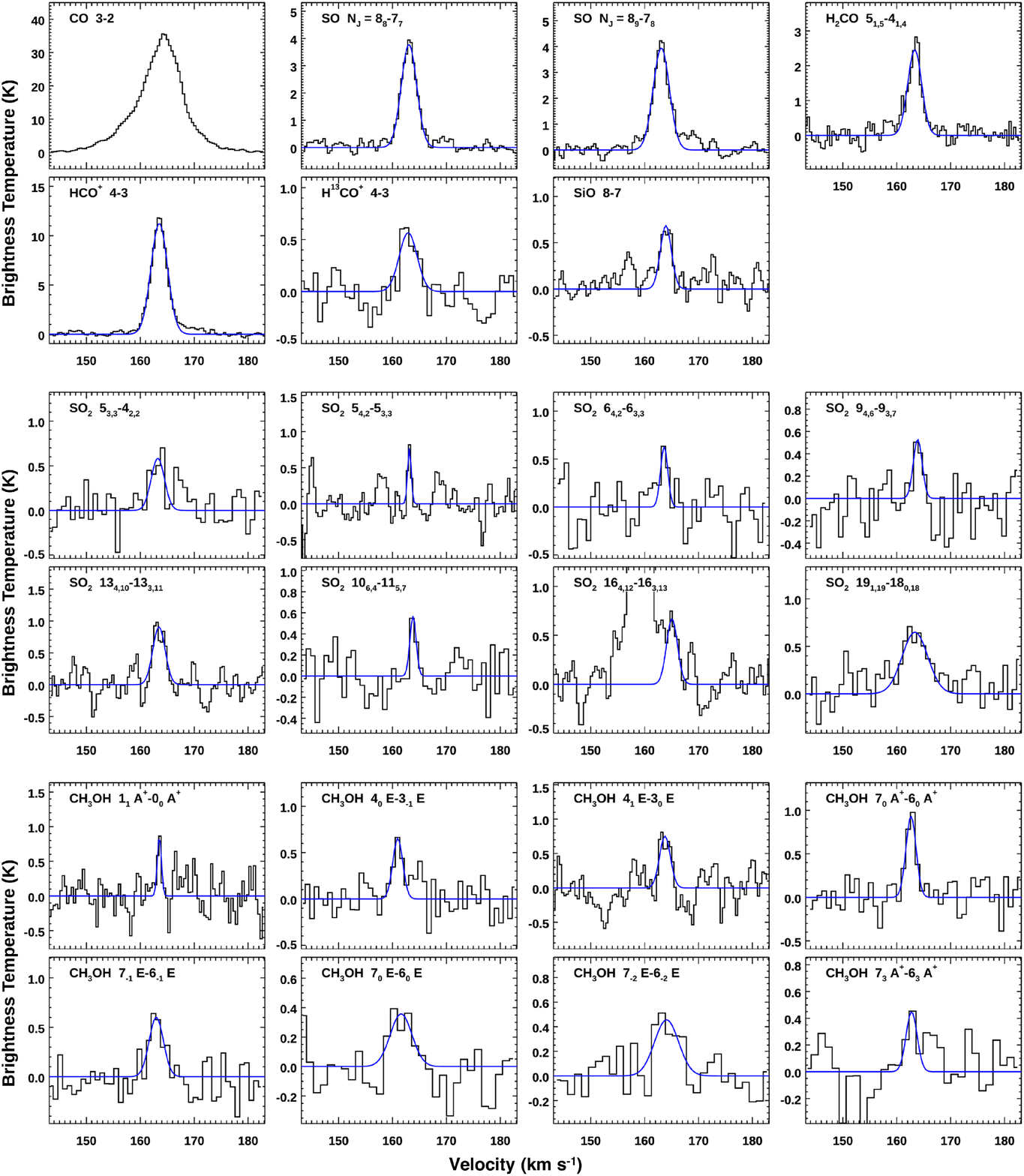} 
\caption{
ALMA spectra of molecular emission lines detected in S07. 
The blue lines represent fitted Gaussian profiles. 
For SO$_2$ and CH$_3$OH, the spectra are sorted in ascending order of the upper state energy (the emission line with the lowest upper state energy is shown in the upper left panel and that with the highest energy is in the lower right panel). 
The adjacent stronger line seen in SO$_2$(16$_{4,12}$--16$_{3,13}$) was subtracted before fitting the main line. 
}
\label{line1}
\end{center}
\end{figure*}

\begin{figure*}[tp]
\begin{center}
\includegraphics[width=17.0cm]{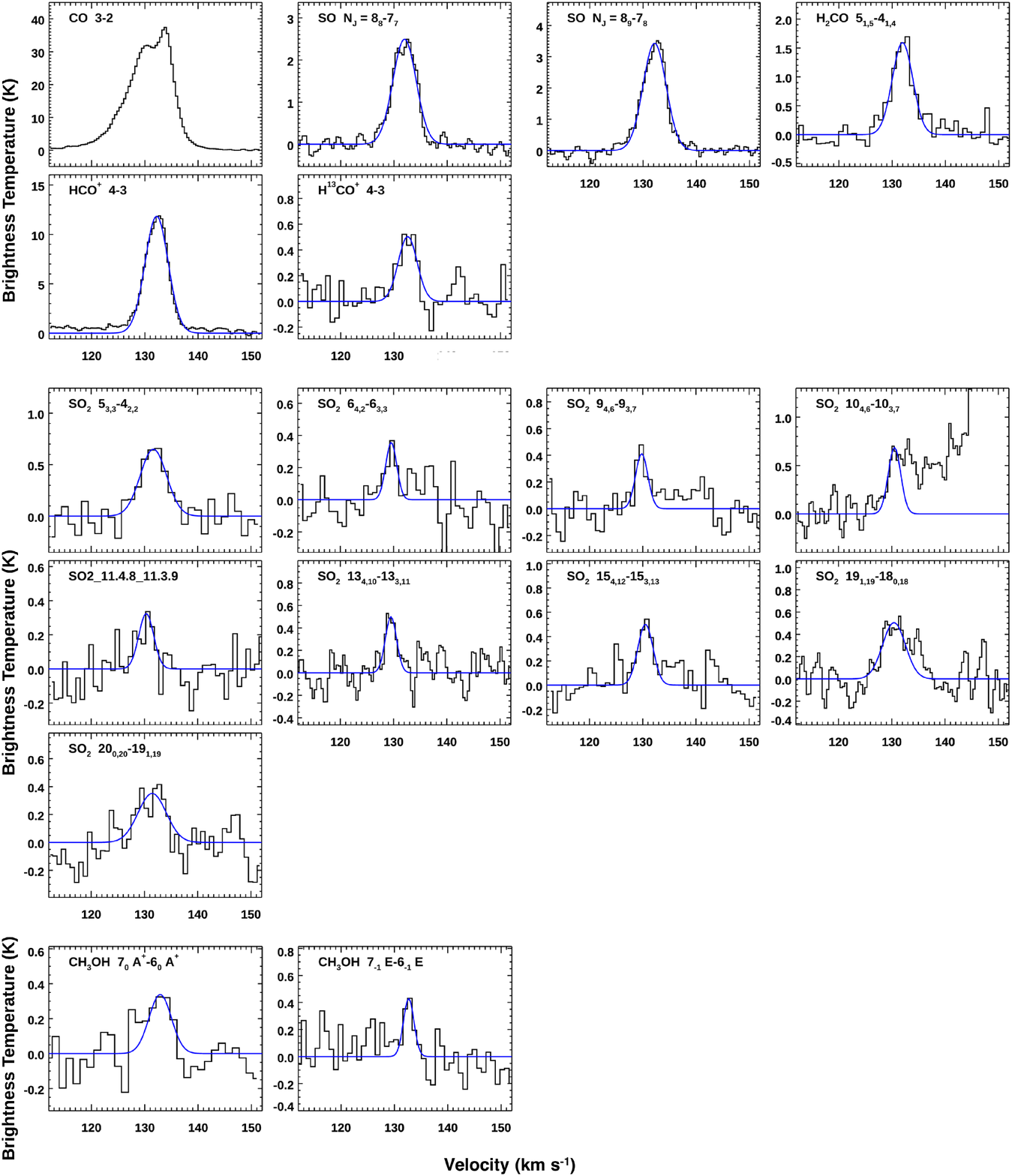}
\caption{
Same as in Figure \ref{line1} but for S09. 
The adjacent stronger line seen in SO$_2$(10$_{4,6}$--10$_{3,7}$) was subtracted before fitting the main line. 
}
\label{line2}
\end{center}
\end{figure*}

Figures \ref{line1}--\ref{line2} show the results of the spectral line fitting (see Section \ref{sec_spc} for details). 

\clearpage

\subsection{Measured line parameters} \label{sec_app_line}
Tables \ref{tab_lines1}--\ref{tab_lines2} summarize line parameters measured by spectral fitting (see Section \ref{sec_spc} for details). 
$T_{br}$, $\Delta$$V$, $\int T_{br} dV$, and $V_{LSR}$ indicate the peak brightness temperature, the FWHM, the integrated intensity, and the systemic velocity. 
For CO(3--2), since it shows a clear non-Gaussian component, the integrated intensity is calculated by directly integrating the spectrum over the frequency region of emission. 
The tables also contain the upper limits on important non-detection lines. 
The tabulated uncertainties and upper limits are of 2$\sigma$ level and do not include systematic errors due to continuum subtraction. 
Upper limits on integrated intensities are estimated assuming $\Delta$$V$ = 4 km s$^{-1}$. 
Note that the systemic velocities of S07 and S09 mentioned in the text is based on the SO($N_J$ = 8$_{9}$--7$_{8}$) line. 

\startlongtable
\begin{deluxetable*}{ l c c c c c c c c c }
\tablecaption{Line Parameters of S07 \label{tab_lines1}}
\tablewidth{0pt}
\tabletypesize{\scriptsize} 
\tablehead{
\colhead{Molecule}                       & \colhead{Transition}                                                             &       \colhead{$E_{u}$} &       \colhead{Frequency} &        \colhead{$T_{br}$} &     \colhead{$\Delta$$V$} &     \colhead{$\int T_{br} dV$} &       \colhead{$V_{LSR}$} &        \colhead{RMS} &       \colhead{Note} \\
\colhead{ }                              & \colhead{ }                                                                      &        \colhead{(K)} &           \colhead{(GHz)} &             \colhead{(K)} &          \colhead{(km/s)} &             \colhead{(K km/s)} &          \colhead{(km/s)} &        \colhead{(K)} &           \colhead{}
}
\startdata
 CO                                      &  3--2                                                                                                                    &   33       & 345.79599       &  35.58 $\pm$   0.39  &   $\sim$10          & 336.3 $\pm$ 0.4    & 164.0           & 0.20 &    (1) \\
 HCO$^+$                                 &  4--3                                                                                                                    &   43       & 356.73422       &  11.25 $\pm$   0.08  &    3.6          & 43.2 $\pm$ 0.7     & 163.5           & 0.24 &    \nodata \\
 H$^{13}$CO$^+$                          &  4--3                                                                                                                    &   42       & 346.99834       &   0.57 $\pm$   0.05  &    3.9          & 2.3 $\pm$ 0.5      & 162.9           & 0.13 &    \nodata \\
 H$_2$CO                                 &  5$_{1,5}$--4$_{1,4}$                                                                                                    &   62       & 351.76864       &   2.46 $\pm$   0.07  &    3.0          & 8.0 $\pm$ 0.5      & 163.3           & 0.24 &    \nodata \\
 SiO                                     &  8--7                                                                                                                    &   75       & 347.33058       &   0.68 $\pm$   0.05  &    2.4          & 1.8 $\pm$ 0.3      & 163.9           & 0.18 &    \nodata \\
 SO                                      &  $N_J$ = 8$_{8}$--7$_{7}$                                                                                                &   87       & 344.31061       &   3.78 $\pm$   0.06  &    3.4          & 13.6 $\pm$ 0.5     & 163.1           & 0.20 &    \nodata \\
 SO                                      &  $N_J$ = 8$_{9}$--7$_{8}$                                                                                                &   79       & 346.52848       &   3.95 $\pm$   0.05  &    3.6          & 14.9 $\pm$ 0.5     & 163.1           & 0.18 &    \nodata \\
 SO$_2$                                  &  20$_{1,19}$--19$_{2,18}$                                                                                                &  199       & 338.61181       &             $<$0.30  &    \nodata      &             $<$1.3   &    \nodata      & 0.15 &    \nodata \\
 SO$_2$                                  &  16$_{4,12}$--16$_{3,13}$                                                                                                &  164       & 346.52388       &   0.67 $\pm$   0.05  &    2.4          & 1.7 $\pm$ 0.3      & 165.1           & 0.18 &    (2) \\
 SO$_2$                                  &  19$_{1,19}$--18$_{0,18}$                                                                                                &  168       & 346.65217       &   0.65 $\pm$   0.05  &    5.6          & 3.9 $\pm$ 0.7      & 163.4           & 0.13 &    \nodata \\
 SO$_2$                                  &  10$_{6,4}$--11$_{5,7}$                                                                                                  &  139       & 350.86276       &   0.57 $\pm$   0.07  &    1.4          & 0.8 $\pm$ 0.3      & 163.8           & 0.17 &    \nodata \\
 SO$_2$                                  &  5$_{3,3}$--4$_{2,2}$                                                                                                    &   36       & 351.25722       &   0.58 $\pm$   0.06  &    2.9          & 1.8 $\pm$ 0.5      & 163.2           & 0.17 &    \nodata \\
 SO$_2$                                  &  14$_{4,10}$--14$_{3,11}$                                                                                                &  136       & 351.87387       &         $<$0.48  &    \nodata      &             $<$2.1   &    \nodata      & 0.24 &    \nodata \\
 SO$_2$                                  &  10$_{4,6}$--10$_{3,7}$                                                                                                  &   90       & 356.75519       &             $<$0.34  &    \nodata      &             $<$1.4   &    \nodata      & 0.17 &    \nodata \\
 SO$_2$                                  &  13$_{4,10}$--13$_{3,11}$                                                                                                &  123       & 357.16539       &   0.91 $\pm$   0.06  &    2.7          & 2.6 $\pm$ 0.5      & 163.5           & 0.24 &    \nodata \\
 SO$_2$                                  &  15$_{4,12}$--15$_{3,13}$                                                                                                &  150       & 357.24119       &      $<$0.48  &    \nodata      &             $<$2.0   &    \nodata      & 0.24 &    \nodata \\
 SO$_2$                                  &  9$_{4,6}$--9$_{3,7}$                                                                                                    &   81       & 357.67182       &   0.53 $\pm$   0.07  &    1.9          & 1.1 $\pm$ 0.4      & 163.9           & 0.17 &    \nodata \\
 SO$_2$                                  &  8$_{4,4}$--8$_{3,5}$                                                                                                    &   72       & 357.58145       &             $<$0.50  &    \nodata      &             $<$2.1   &    \nodata      & 0.17 &    \nodata \\
 SO$_2$                                  &  7$_{4,4}$--7$_{3,5}$                                                                                                    &   65       & 357.89244       &             $<$0.34  &    \nodata      &             $<$1.4   &    \nodata      & 0.17 &    \nodata \\
 SO$_2$                                  &  6$_{4,2}$--6$_{3,3}$                                                                                                    &   59       & 357.92585       &   0.63 $\pm$   0.06  &    1.6          & 1.1 $\pm$ 0.3      & 163.6           & 0.17 &    \nodata \\
 SO$_2$                                  &  17$_{4,14}$--17$_{3,15}$                                                                                                &  180       & 357.96290       &             $<$0.34  &    \nodata      &             $<$1.4   &    \nodata      & 0.17 &    \nodata \\
 SO$_2$                                  &  5$_{4,2}$--5$_{3,3}$                                                                                                    &   53       & 358.01315       &   0.77 $\pm$   0.08  &    0.8          & 0.7 $\pm$ 0.1      & 163.1           & 0.24 &    \nodata \\
 SO$_2$                                  &  20$_{0,20}$--19$_{1,19}$                                                                                                &  185       & 358.21563       &             $<$0.39  &    \nodata      &             $<$1.7   &    \nodata      & 0.20 &    \nodata \\
 CH$_3$OH                                &  7$_{0}$ E--6$_{0}$ E                                                                                                    &   78       & 338.12449       &   0.36 $\pm$   0.06  &    4.8          & 1.8 $\pm$ 0.7      & 161.6           & 0.12 &    \nodata \\
 CH$_3$OH                                &  7$_{-1}$ E--6$_{-1}$ E                                                                                                  &   71       & 338.34459       &   0.60 $\pm$   0.06  &    3.1          & 2.0 $\pm$ 0.4      & 163.0           & 0.15 &    \nodata \\
 CH$_3$OH                                &  7$_{0}$ A$^+$--6$_{0}$ A$^+$                                                                                            &   65       & 338.40870       &   0.93 $\pm$   0.06  &    2.2          & 2.2 $\pm$ 0.3      & 162.6           & 0.15 &    \nodata \\
 CH$_3$OH                                &  7$_{2}$ A$^-$--6$_{2}$ A$^-$                                                                                            &  103       & 338.51285       &             $<$0.24  &    \nodata      &             $<$1.0   &    \nodata      & 0.12 &    (3) \\
 CH$_3$OH                                &  7$_{3}$ A$^+$--6$_{3}$ A$^+$                                                                                            &  115       & 338.54083       &   0.45 $\pm$   0.06  &    2.4          & 1.2 $\pm$ 0.4      & 162.7           & 0.12 &    (4) \\
 CH$_3$OH                                &  7$_{-3}$ E--6$_{-3}$ E                                                                                                  &  128       & 338.55996       &             $<$0.24  &    \nodata      &             $<$1.0   &    \nodata      & 0.12 &    \nodata \\
 CH$_3$OH                                &  7$_{3}$ E--6$_{3}$ E                                                                                                    &  113       & 338.58322       &             $<$0.24  &    \nodata      &             $<$1.0   &    \nodata      & 0.12 &    \nodata \\
 CH$_3$OH                                &  7$_{1}$ E--6$_{1}$ E                                                                                                    &   86       & 338.61494       &             $<$0.50  &    \nodata      &             $<$2.1   &    \nodata      & 0.15 &    \nodata \\
 CH$_3$OH                                &  7$_{2}$ A$^+$--6$_{2}$ A$^+$                                                                                            &  103       & 338.63980       &             $<$0.24  &    \nodata      &             $<$1.0   &    \nodata      & 0.12 &    \nodata \\
 CH$_3$OH                                &  7$_{-2}$ E--6$_{-2}$ E                                                                                                  &   91       & 338.72290       &   0.46 $\pm$   0.06  &    5.3          & 2.6 $\pm$ 0.7      & 164.1           & 0.12 &    (4) \\
 CH$_3$OH                                &  5$_{4}$ A$^-$--6$_{3}$ A$^-$                                                                                            &  115       & 346.20272       &             $<$0.26  &    \nodata      &             $<$1.1   &    \nodata      & 0.13 &    (4) \\
 CH$_3$OH                                &  4$_{0}$ E--3$_{-1}$ E                                                                                                   &   36       & 350.68766       &   0.65 $\pm$   0.07  &    2.2          & 1.5 $\pm$ 0.4      & 161.0           & 0.17 &    \nodata \\
 CH$_3$OH                                &  1$_{1}$ A$^+$--0$_{0}$ A$^+$                                                                                            &   17       & 350.90510       &   0.87 $\pm$   0.09  &    0.9          & 0.8 $\pm$ 0.2      & 163.5           & 0.24 &    \nodata \\
 CH$_3$OH                                &  4$_{1}$ E--3$_{0}$ E                                                                                                    &   44       & 358.60580       &   0.75 $\pm$   0.07  &    2.4          & 1.9 $\pm$ 0.4      & 163.8           & 0.28 &    \nodata \\
\enddata
\tablecomments{
(1) $\int T_{br} dV$ is calculated by directly integrating the spectrum from 145 km s$^{-1}$ to 180 km s$^{-1}$. 
(2) Partial blend with SO($N_J$ = 8$_{9}$--7$_{8}$). 
(3) Blend of three CH$_3$OH lines with similar spectroscopic constants.
(4) Blend of two CH$_3$OH lines with similar spectroscopic constants. 
}
\end{deluxetable*}

\startlongtable
\begin{deluxetable*}{ l c c c c c c c c c }
\tablecaption{Line Parameters of S09 \label{tab_lines2}}
\tablewidth{0pt}
\tabletypesize{\scriptsize} 
\tablehead{
\colhead{Molecule}                       & \colhead{Transition}                                                             &       \colhead{$E_{u}$} &       \colhead{Frequency} &        \colhead{$T_{br}$} &     \colhead{$\Delta$$V$} &     \colhead{$\int T_{br} dV$} &       \colhead{$V_{LSR}$} &        \colhead{RMS} &       \colhead{Note} \\
\colhead{ }                              & \colhead{ }                                                                      &        \colhead{(K)} &           \colhead{(GHz)} &             \colhead{(K)} &          \colhead{(km/s)} &             \colhead{(K km/s)} &          \colhead{(km/s)} &        \colhead{(K)} &           \colhead{}
}
\startdata
 CO                                      &  3--2                                                                                                                    &   33       & 345.79599       &  37.40 $\pm$   0.27  &   $\sim$10          & 361.4 $\pm$ 0.3    & 131.3           & 0.13 &    (1) \\
 HCO$^+$                                 &  4--3                                                                                                                    &   43       & 356.73422       &  11.84 $\pm$   0.05  &    5.1          & 64.47 $\pm$ 0.56     & 132.2           & 0.14 &    \nodata \\
 H$^{13}$CO$^+$                          &  4--3                                                                                                                    &   42       & 346.99834       &   0.50 $\pm$   0.04  &    4.2          & 2.25 $\pm$ 0.37      & 132.6           & 0.09 &    \nodata \\
 H$_2$CO                                 &  5$_{1,5}$--4$_{1,4}$                                                                                                    &   62       & 351.76864       &   1.63 $\pm$   0.06  &    4.4          & 7.66 $\pm$ 0.58      & 131.9           & 0.20 &    \nodata \\
 SiO                                     &  8--7                                                                                                                    &   75       & 347.33058       &             $<$0.18  &    \nodata      &             $<$0.8   &    \nodata      & 0.09 &    \nodata \\
 SO                                      &  $N_J$ = 8$_{8}$--7$_{7}$                                                                                                &   87       & 344.31061       &   2.50 $\pm$   0.04  &    5.2          & 13.70 $\pm$ 0.49     & 132.0           & 0.13 &    \nodata \\
 SO                                      &  $N_J$ = 8$_{9}$--7$_{8}$                                                                                                &   79       & 346.52848       &   3.43 $\pm$   0.04  &    5.1          & 18.70 $\pm$ 0.49     & 132.2           & 0.13 &    \nodata \\
 SO$_2$                                  &  20$_{1,19}$--19$_{2,18}$                                                                                                &  199       & 338.61181       &             $<$0.50  &    \nodata      &             $<$2.1   &    \nodata      & 0.12 &    \nodata \\
 SO$_2$                                  &  16$_{4,12}$--16$_{3,13}$                                                                                                &  164       & 346.52388       &             $<$0.80  &    \nodata      &             $<$3.4   &    \nodata      & 0.13 &    (2) \\
 SO$_2$                                  &  19$_{1,19}$--18$_{0,18}$                                                                                                &  168       & 346.65217       &   0.51 $\pm$   0.03  &    5.0          & 2.72 $\pm$ 0.52      & 130.4           & 0.13 &    \nodata \\
 SO$_2$                                  &  10$_{6,4}$--11$_{5,7}$                                                                                                  &  139       & 350.86276       &             $<$0.27  &    \nodata      &             $<$1.4   &    \nodata      & 0.14 &    \nodata \\
 SO$_2$                                  &  5$_{3,3}$--4$_{2,2}$                                                                                                    &   36       & 351.25722       &   0.65 $\pm$   0.05  &    5.9          & 4.05 $\pm$ 0.73      & 131.7           & 0.11 &    \nodata \\
 SO$_2$                                  &  14$_{4,10}$--14$_{3,11}$                                                                                                &  136       & 351.87387       &              $<$0.39  &    \nodata      &             $<$1.7   &    \nodata      & 0.20 &    \nodata \\
 SO$_2$                                  &  10$_{4,6}$--10$_{3,7}$                                                                                                  &   90       & 356.75519       &   0.69 $\pm$   0.04  &    2.8          & 2.02 $\pm$ 0.68      & 130.51           & 0.15 &    (3) \\
 SO$_2$                                  &  13$_{4,10}$--13$_{3,11}$                                                                                                &  123       & 357.16539       &   0.50 $\pm$   0.04  &    2.5          & 1.33 $\pm$ 0.26      & 129.5           & 0.14 &    \nodata \\
 SO$_2$                                  &  15$_{4,12}$--15$_{3,13}$                                                                                                &  150       & 357.24119       &   0.50 $\pm$   0.04  &    3.4          & 1.80 $\pm$ 0.32      & 130.5           & 0.10 &    \nodata \\
 SO$_2$                                  &  11$_{4,8}$--11$_{3,9}$                                                                                                  &  100       & 357.38758       &   0.33 $\pm$   0.04  &    3.3          & 1.14 $\pm$ 0.39      & 130.3           & 0.10 &    \nodata \\
 SO$_2$                                  &  9$_{4,6}$--9$_{3,7}$                                                                                                    &   81       & 357.67182       &   0.41 $\pm$   0.04  &    2.7          & 1.20 $\pm$ 0.26      & 129.8           & 0.10 &    \nodata \\
 SO$_2$                                  &  8$_{4,4}$--8$_{3,5}$                                                                                                    &   72       & 357.58145       &             $<$0.29  &    \nodata      &             $<$1.2   &    \nodata      & 0.14 &    \nodata \\
 SO$_2$                                  &  7$_{4,4}$--7$_{3,5}$                                                                                                    &   65       & 357.89244       &             $<$0.29  &    \nodata      &             $<$1.2   &    \nodata      & 0.14 &    \nodata \\
 SO$_2$                                  &  6$_{4,2}$--6$_{3,3}$                                                                                                    &   59       & 357.92585       &   0.36 $\pm$   0.04  &    2.5          & 0.93 $\pm$ 0.29      & 129.6           & 0.10 &    \nodata \\
 SO$_2$                                  &  17$_{4,14}$--17$_{3,15}$                                                                                                &  180       & 357.96290       &        $<$0.29  &    \nodata      &             $<$1.2   &    \nodata      & 0.14 &    \nodata \\
 SO$_2$                                  &  5$_{4,2}$--5$_{3,3}$                                                                                                    &   53       & 358.01315       &             $<$0.29  &    \nodata      &             $<$1.2   &    \nodata      & 0.14 &    \nodata \\
 SO$_2$                                  &  4$_{4,0}$--4$_{3,1}$                                                                                                    &   48       & 358.03789       &             $<$0.20  &    \nodata      &             $<$0.9   &    \nodata      & 0.10 &    \nodata \\
 SO$_2$                                  &  20$_{0,20}$--19$_{1,19}$                                                                                                &  185       & 358.21563       &   0.35 $\pm$   0.04  &    6.2          & 2.32 $\pm$ 0.70      & 131.4           & 0.12 &    \nodata \\
 CH$_3$OH                                &  7$_{0}$ E--6$_{0}$ E                                                                                                    &   78       & 338.12449       &             $<$0.25  &    \nodata      &             $<$1.3   &    \nodata      & 0.12 &    \nodata \\
 CH$_3$OH                                &  7$_{-1}$ E--6$_{-1}$ E                                                                                                  &   71       & 338.34459       &   0.43 $\pm$   0.05  &    2.3          & 1.06 $\pm$ 0.31      & 132.7           & 0.12 &    \nodata \\
 CH$_3$OH                                &  7$_{0}$ A$^+$--6$_{0}$ A$^+$                                                                                            &   65       & 338.40870       &   0.34 $\pm$   0.04  &    4.9          & 1.78 $\pm$ 0.60      & 132.9           & 0.10 &    \nodata \\
 CH$_3$OH                                &  7$_{1}$ E--6$_{1}$ E                                                                                                    &   86       & 338.61494       &             $<$0.25  &    \nodata      &             $<$1.3   &    \nodata      & 0.12 &    \nodata \\
 CH$_3$OH                                &  7$_{2}$ A$^+$--6$_{2}$ A$^+$                                                                                            &  103       & 338.63980       &             $<$0.25  &    \nodata      &             $<$1.3   &    \nodata      & 0.12 &    \nodata \\
 CH$_3$OH                                &  4$_{0}$ E--3$_{-1}$ E                                                                                                   &   36       & 350.68766       &             $<$0.22  &    \nodata      &             $<$0.9   &    \nodata      & 0.11 &    \nodata \\
 CH$_3$OH                                &  1$_{1}$ A$^+$--0$_{0}$ A$^+$                                                                                            &   17       & 350.90510       &             $<$0.20  &    \nodata      &             $<$0.9   &    \nodata      & 0.10 &    \nodata \\
 CH$_3$OH                                &  4$_{1}$ E--3$_{0}$ E                                                                                                    &   44       & 358.60580       &             $<$0.25  &    \nodata      &             $<$1.0   &    \nodata      & 0.12 &    \nodata \\
\enddata
\tablecomments{
(1) $\int T_{br} dV$ is calculated by directly integrating the spectrum from 114 km s$^{-1}$ to 145 km s$^{-1}$. 
(2) Partial blend with SO($N_J$ = 8$_{9}$--7$_{8}$). 
(3) Partial blend with HCO$^+$(4--3). 
}
\end{deluxetable*}

\clearpage

\section{Derivation of column densities and temperatures} \label{sec_app_calc}
\subsection{Rotation diagram analysis} \label{sec_app_rd}
We use the following formulae based on the standard treatment of the rotation diagram method \citep[e.g., ][]{Sut95}: 
\begin{equation}
\log \left(\frac{ N_{u} }{ g_{u} } \right) = - \left(\frac {\log e}{T_{\mathrm{rot}}} \right) \left(\frac{E_{u}}{k} \right) + \log \left(\frac{N}{Q(T_{\mathrm{rot}})} \right),  \label{Eq_rd1}
\end{equation}
where 
\begin{equation}
\frac{ N_{u} }{ g_{u} } = \frac{ 3 k \int T_{\mathrm{b}} dV }{ 8 \pi^{3} \nu S \mu^{2} }, \label{Eq_rd2} \\ 
\end{equation}
and $N_{u}$ is a column density of molecules in the upper energy level, $g_{u}$ is the degeneracy of the upper level, $k$ is the Boltzmann constant, $\int T_{\mathrm{b}} dV$ is the integrated intensity estimated from the observations, $\nu$ is the transition frequency, $S$ is the line strength, $\mu$ is the dipole moment, $T_{\mathrm{rot}}$ is the rotational temperature, $E_{u}$ is the upper state energy, $N$ is the total column density, and $Q(T_{\mathrm{rot}})$ is the partition function at $T_{\mathrm{rot}}$. 
We assume an optically thin condition and the local thermodynamic equilibrium (LTE). 
All the spectroscopic parameters required in the analysis are extracted from the CDMS database.

\subsection{Derivation of the H$_2$ column density from the dust continuum} \label{sec_app_h2} 
We use the following equation to calculate $N_{\mathrm{H_2}}$ based on the standard treatment of optically thin dust emission: 
\begin{equation}
N_{\mathrm{H_2}} = \frac{F_{\nu} / \Omega}{2 \kappa_{\nu} B_{\nu}(T_{d}) Z \mu m_{\mathrm{H}}} \label{Eq_h2}, 
\end{equation}
where $F_{\nu}/\Omega$ is the continuum flux density per beam solid angle as estimated from the observations, $\kappa_{\nu}$ is the mass absorption coefficient of dust grains coated by thin ice mantles taken from \citet{Oss94} and we use 1.94 cm$^2$ g$^{-1}$ at 850 $\mu$m, $T_{d}$ is the dust temperature and $B_{\nu}(T_{d})$ is the Planck function, $Z$ is the dust-to-gas mass ratio, $\mu$ is the mean atomic mass per hydrogen \citep[1.41, according to][]{Cox00}, and $m_{\mathrm{H}}$ is the hydrogen mass. 
We use the dust-to-gas mass ratio of 0.0016, which is obtained by scaling the Galactic value of 0.008 by the metallicity of the SMC. 
The assumed dust temperature is described in the main text.

\section{Molecular outflows traced by CO} \label{sec_app_outflow}
\begin{figure*}[thp!]
\begin{center}
\includegraphics[width=12.0cm]{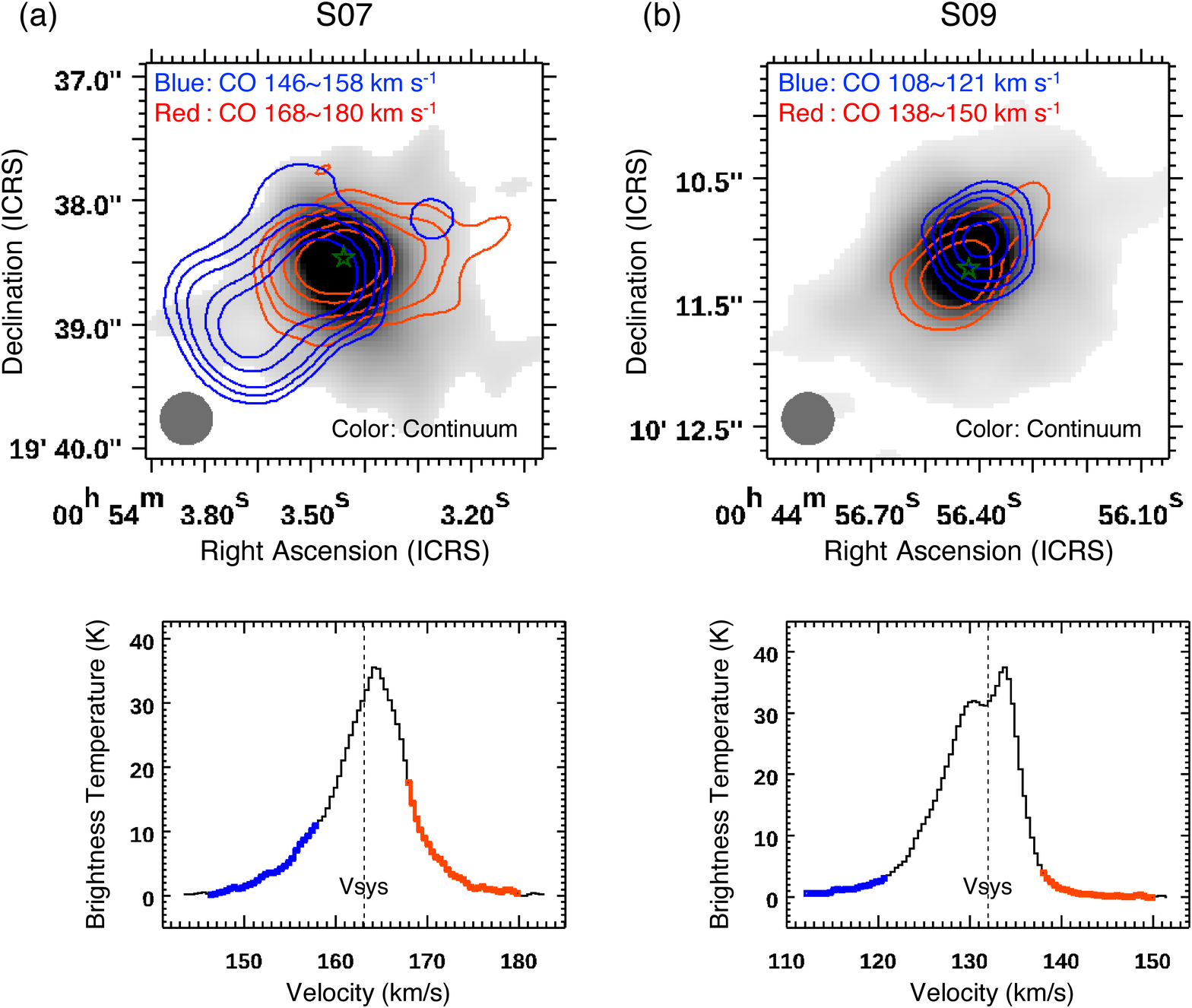}
\caption{
CO(3-2) outflows detected in S07 (a) and S09 (b). 
The upper panels show the spatial distribution of high-velocity CO gas. 
Blue contours represent the blue-shifted component (integrated over 146--158 km $s^{-1}$ for S07 and 108--121 km $s^{-1}$ for S09), 
while red contours represent the red-shifted component (integrated over 168--180 km $s^{-1}$ for S07 and 138--150 km $s^{-1}$ for S09). 
Background is the 850 $\mu$m continuum flux. 
The green stars represent the position of the protostar inferred from the continuum peak position. 
The synthesized beam size is shown by the gray filled circle in each panel. 
The lower panels show the CO(3-2) spectra extracted at the hot core position as in Figures \ref{line1}--\ref{line2}. 
The velocity ranges of the blue- and red-shifted high-velocity components are indicated by blue and red, respectively. 
The dotted lines indicate a systemic velocity of the source estimated from the SO($N_J$ = 8$_{8}$--7$_{7}$) line. 
}
\label{outflow}
\end{center}
\end{figure*}

Protostellar molecular outflows are detected in S07 and S09 as high-velocity CO(3-2) components (Fig.\ref{outflow}). 
The presence of shocked H$_2$ gas associated with the protostar is reported in \citet{War17} for both sources based on high-spatial resolution near-infrared integral field spectroscopy. 
The molecular component of outflows in S07 is recently reported by \citet{Tok22} with ALMA observations of the CO(3-2) line. 
The outflows detected in S09 is the second example of CO-traced protostellar outflows found in the SMC. 
A detailed analysis of protostellar outflows for a sample of LMC/SMC YSOs will be reported in a future paper based on the MAGOS data (Tanaka et al. in prep.).

\end{document}